\documentclass[pre,twocolumn,showpacs,preprintnumbers,amsmath,amssymb]{revtex4}

\usepackage{graphicx}
\usepackage{dcolumn}
\usepackage{bm}
\usepackage{subfigure}
\usepackage{hyperref}

\begin{document}

\title{Quantum information entropies of the eigenstates of the Morse potential}
\author{Ekrem Ayd\i ner}
\email{ekrem.aydiner@deu.edu.tr}
\author{Cenk Orta}
\affiliation{Dokuz Eyl\"{u}l University, Department of Physics
35160 \.{I}zmir, Turkey}
\author{Ramazan Sever}
\affiliation{Middle East Technical University, Department of
Physics 06531 Ankara, Turkey}


\begin{abstract}
The position and momentum space information entropies for the Morse
potential are numerically obtained for different strengths of the
potential. It is found to satisfy the bound obtained by Beckner,
Bialynicki-Birula, and Mycielski. Interesting features of the
entropy densities are graphically demonstrated.
\end{abstract}

\maketitle

\section{Introduction}
In quantum mechanics two non-commuting observables cannot be
simultaneously measured with arbitrary precision. That is, there is
an irreducible lower bound on the uncertainty in the result of a
simultaneous measurement of non-commuting observables. This fact,
often called the Heisenberg uncertainty principle \cite{Heisenberg}.
The uncertainty principle specified for a given pairs of observables
finds its mathematical manifestation as the uncertainty relations.
The first rigorous derivation of the uncertainty relation from the
quantum mechanical formalism applied for the basic non-commuting
observables is due to Kennard \cite{Kennard}.

In accordance with the present understanding the quantum system is
described by a complex function $\psi \left( x,t\right)$, which is
linked with the function of the probability density of finding a
particle at the position $x$ at the time $t$ by the equation $\rho
_{x}\left( x,t\right) =\left\vert \psi \left( x,t\right) \right\vert
^{2}$. On the other hand, the corresponding Fourier transform $\phi
\left( x,t\right)$ is connected with the probability of finding a
particle with momentum $p$ at the time $t$ by the equation $\rho
_{p}\left( p,t\right) =\left\vert \phi \left( p,t\right) \right\vert
^{2}$.

For the mathematical expression of the uncertainty principle we must
more precisely define how to measure the sharpness of the
probability density function of an observable. In the mathematical
formulation of the uncertainty principle we consider two Hermitian
operators $\widehat{A}$ and $\widehat{B}$ which represent physical
observables $A$ and $B$ in a finite $N$-dimensional Hilbert space.
Let $\left\{\left\vert a_{j}\right\rangle \right\}$ and $\left\{
\left\vert b_{k}\right\rangle \right\}$, $j,k=1,2,...,N$, be
corresponding complete sets of normalized eigenvectors. The
probability distribution of the observables $A$ and $B$, $P=\left\{
p_{1},p_{2},...,p_{N}\right\}$ and $Q=\left\{
q_{1},q_{2},...,q_{N}\right\}$ described by the wave-function $\psi$
are given by the equations
\begin{equation}
p_{j}=\left\vert \left\langle a_{j}\mid \psi \right\rangle
\right\vert ^{2}, \ \ \ q_{k}=\left\vert \left\langle b_{k}\mid \psi
\right\rangle \right\vert ^{2}
\end{equation}
respectively. The principle of uncertainty says that if $A$ and $B$
are non-commuting observables, then their probability distributions
cannot be both arbitrarily peaked. In other words, the uncertainty
principle states that two complementary observables cannot have the
same eigenfunctions.

The formulation of the uncertainty principle by means of dispersions
of non-commuting observables is usually given in the form of the
Robertson relation \cite{Robertson}:
\begin{equation}
\Delta A\Delta B\geq \frac{1}{2}\vert \langle \psi \vert [
\widehat{A},\widehat{B}] \vert \psi \rangle \vert
\end{equation}
where $\Delta A$ and $\Delta B$ denote the standard deviation of
distributions (1):
\begin{equation*}
\Delta A=[ \langle A^{2}\rangle - \langle A \rangle ^{2}]^{1/2}, \ \
\Delta B=[ \langle B^{2}\rangle - \langle B \rangle ^{2}]^{1/2}.
\end{equation*}
It has been pointed by many authors (See Refs.
\cite{MaassenUffink,Deutsch,UffinkHilgewoord}) that the Robertson
form of uncertainty relation has two serious shortcomings; (i) the
right-hand side of Eq.\,(2) is not a fixed lower bound, but depends
on the state $\psi$. If the observables $A$ and $B$ is in its
eigenstate then $\left[ A,B\right] =0$ and no restriction on $\Delta
A$ or $\Delta B$ is imposed by the left-hand side of the inequality
(2). (ii) The dispersion may not represent the appropriate measure
for the uncertainty of an observables if its probability
distribution exhibits some sharp distant peaks.

Therefore, to improve on this situation entropic uncertainty
relations have been proposed which rely on the Shannon entropy
\cite{Shannon} as a measure of uncertainty. A much more satisfactory
measure of quantum uncertainty is given by the information entropy
of given probability distribution. An entropic uncertainty relation
represents the sum of the entropies of two non-commuting observables
$A$ and $B$
\begin{equation}
S_{A}+S_{B}\geq S_{AB}
\end{equation}
where $S_{A}$ and $S_{B}$ denote information entropies of
observables $A$ and $B$, respectively. The $S_{AB}$ is a positive
constant which represent the lower bound of the right-hand side of
the inequality of (3). The information uncertainty relation (3) were
first conjectured by Everett \cite{Everett} in the context of
\textit{many worlds interpretation} of quantum mechanics and
Hirschman \cite{Hirschmann} in 1957, and proved by Bialynicki-Birula
and Mycielski \cite{Bialynicki1}, and independently by Beckner
\cite{Beckner}.

For the continuous observables for example $x$ and $p$ which are
described by the wave-functions $\psi(x)$ and $\phi(p)$, position
space information entropy $S_{x}$ and momentum space information
entropy $S_{p}$ are defined by
\begin{equation}
S_{x}=-\int_{-\infty }^{\infty }\left\vert \psi \left( x\right)
\right\vert ^{2}\ln \left\vert \psi \left( x\right) \right\vert
^{2}dx
\end{equation}
\begin{equation}
S_{p}=-\int_{-\infty }^{\infty }\left\vert \phi \left( p\right)
\right\vert ^{2}\ln \left\vert \phi \left( p\right) \right\vert
^{2}dp
\end{equation}
respectively. For position $x$ and momentum $p$ the inequality (3)
for $N$-dimensional Hilbert space reads
\begin{equation}
S_{x}+S_{p}\geq N(1+ln\pi)
\end{equation}
where $N(1+ln\pi)$ is the lower bound for the inequality (3). For
$N=1$, the entropy sum is bounded from below by the value $2.1447...
\ .$

Note that these entropies have been used for numerous practical
purposes such as, for example, to measure the squeezing of quantum
fluctuation \cite{Orlowski} and to reconstruct the charge and
momentum densities of atomic and molecular systems
\cite{Galindo,Angulo} by means of maximum-entropy procedures.

\section{The Morse Potential}
In present study we deal with position and momentum information
entropies of the one-dimensional Morse potential as an example, and
will discuss whether it satisfy the entropic uncertainty relation
(6) or not. The Morse potential, introduced in the 1930's
\cite{Morse}, is one of a class of potential \cite{PoschlTeller} for
which solutions of the Schr\"{o}dinger equation are known. The Morse
potential is widely used in the literature as a model for bound
states, such as the vibrational states of molecules. For the
applications the potential is defined as a function of the variable
$x$ which ranges from $-\infty$ to $+\infty$, and is given in terms
of two parameters $D$ and $\alpha$ by
\begin{equation}
V(x)=De^{-\alpha x}(e^{-\alpha x}-2)
\end{equation}
where $D$ is the dissociation energy, $D>0$ corresponds to its
depth, and $\alpha$ is related to the range of the potential and $x$
gives the relative distance from the equilibrium position of the
atoms. At the $x=0$, it has a negative (attractive) minimum of depth
$D$, and it goes smoothly to zero in the limit of large $x$. For $x$
less than $-\ln 2/\alpha$ this potential becomes positive
(repulsive), and as $x$ decrease even further, the potential becomes
increasingly large.

The solution of the Schr\"{o}dinger equation associated with the
potential (7) is given by \cite{Morse}
\begin{equation}
\Psi_{n}^{\lambda}(\xi)=Ne^{-\xi/2}\xi^{s/2}L_n^{s}(\xi)
\end{equation}
where $L_n^{s}(\xi)$ are associated Laguerre functions, the argument
$\xi$ is related to the physical displacement coordinate $x$ by $\xi
= 2 \lambda e^{-\alpha x}$ with $0<\xi <\infty$, and $N$ is
normalization constant:
\begin{equation}
N=\sqrt{\frac{\alpha \left(2\lambda - 2n -1\right) \Gamma \left( n+1\right) }{%
\Gamma \left(2\lambda-n \right) }}
\end{equation}
and $n=0,1,...,\left[ \lambda -1/2\right]$ with $\left[ \rho
\right]$ denoting the largest integer smaller than $\rho$, so that
the total number of bound states is $\left[ \lambda -1/2\right]$.

We note that the $\lambda$ is potential dependent, $s$ is related to
energy $E$ and by definition $\lambda>0$, $s>0$. The parameters
$\lambda$ and $s$ satisfy the constraint condition
\begin{equation}
s+2n=2\lambda -1
\end{equation}
end they are related to the potential $D$ and energy $E$ through
\begin{equation}
\lambda=\sqrt{\frac{2\mu D}{\alpha ^{2}\hbar ^{2}}} \ \ \ \ \ \ s=\sqrt{%
\frac{-8\mu E}{\alpha ^{2}\hbar ^{2}}} \ .
\end{equation}
where $\mu$ is the reduced mass of the molecule. Normalizable states
fulfil $n<s$ and the corresponding eigenvalues (i.e., energy
spectrum) are given by
\begin{equation}
E_n=-\hbar\omega(n+1/2)
\end{equation}
where $w=\hbar \alpha ^{2}/2\mu $.
\begin{figure*}
\includegraphics[width=13.5cm]{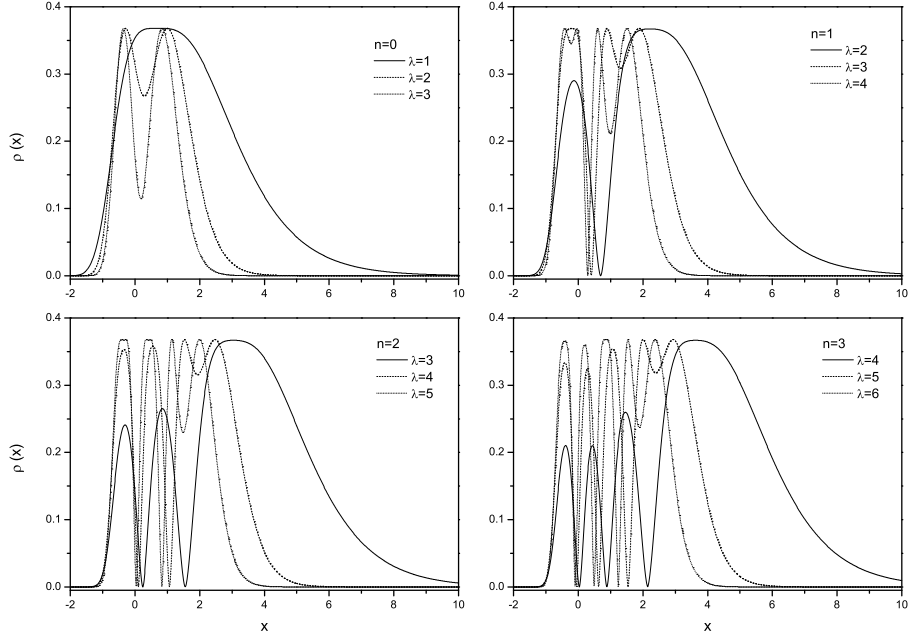}
\caption{Plots of the position space entropy densities of the Morse
potential for (a) $n=0$ and $\lambda=1,2,3$ (b) $n=1$ and
$\lambda=2,3,4$ (c) $n=2$ and $\lambda=3,4,5$ (d) $n=3$ and
$\lambda=4,5,6$, respectively.}
\end{figure*}
\begin{figure*}
\includegraphics[width=13.5cm]{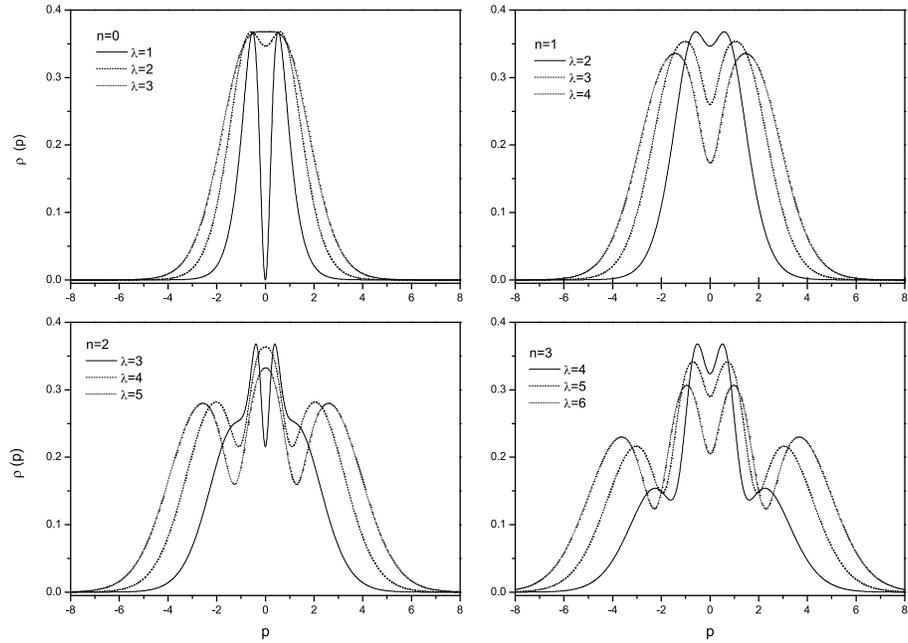}
\caption{Plots of the momentum space entropy densities of the Morse
potential for (a) $n=0$ and $\lambda=1,2,3$ (b) $n=1$ and
$\lambda=2,3,4$ (c) $n=2$ and $\lambda=3,4,5$ (d) $n=3$ and
$\lambda=4,5,6$, respectively.}
\end{figure*}

\section{The Numerical Results}
The position and momentum space information entropies for the
one-dimensional potential can be calculated by using Eqs.\,(4) and
(5), respectively. The analytical derivation of the position and
momentum space information entropies for Morse potential is quite
cumbersome. However, quite recently position and momentum space
information entropies of the Morse have been analytically obtained
by Dehesa et al. for the ground state of the wave-function
\cite{DehesaSorokin}. They also discussed behavior of the ground
state entropy for values of the potential parameter $D$. Therefore,
the present article is devoted to the numerical study of the
information entropies for Morse potential take into account the
parameters of the wave-function (8). Instead evolution of position
and momentum space information entropies, we plot entropy densities
 $\rho(x)=\left\vert \psi \left(
x\right) \right\vert ^{2}\ln \left\vert \psi \left( x\right)
\right\vert ^{2}$ and $\rho(p)=\left\vert \phi \left( p\right)
\right\vert ^{2}\ln \left\vert \phi \left( p\right) \right\vert
^{2}$ for both position and momentum space, respectively. The
entropy densities $\rho(x)$ and $\rho(p)$ provide a measure of
information about localization of the particle in the respective
spaces.

In order to demonstrating the entropy distribution in the well, we
have plotted position and momentum space entropy densities in
Figs.\,1(a)-(d) and Figs.\,2(a)-(d), respectively, for arbitrary $n$
and $\lambda$ values. It seems that in the both figures the number
of minima and their depths depend on $n$ and $\lambda$. However,
entropy distributions exhibit interesting behavior, for example, the
distribution of the position space entropy density has asymmetric
shape as can be seen in Figs.\,1(a)-(d), whereas, the distribution
of the momentum space entropy density in momentum space are quite
symmetric as can be seen in Figs.\,2(a)-(d).

In Table\,1 we have presented the numerical results of information
entropies $S_{x}$, $S_{p}$ and the entropy sum $S_{x}+ S_{p}$ for
arbitrary $n$ and $\lambda$. It is clearly seen from Table\,1 that
BBM inequality is satisfied for the Morse potential. For all $n$
values, as $\lambda$ increases, the entropy sum $S_{x}+ S_{p}$ tends
to be saturated to bound value which is defined by BBM inequality.
Physically, for increasing $\lambda$, the depth of the potential
increases and it increasingly resembles the oscillator potential,
which saturates above inequality. The significance of BBM inequality
is that it presents an irreducible lower bound to the entropy sum.
The conjugate position and momentum space information entropies have
an inverse relationship with each other. A strongly localized
distribution in the position space corresponds to widely delocalized
distribution in the momentum space. With one entropy increasing, the
other entropy decreases but only to the extent that their sum stays
above the stipulated lower bound of $(1+\ln\pi$).
\begin{table}
\caption{\label{tabone}Table for BBM inequality for the Morse
potential for arbitrary $n$ and $\lambda$}
\begin{ruledtabular}
\begin{tabular}{cccccc}
 n&$\lambda$&$S_{x}$&$S_{p}$&
$ S_{x}+S_{p}$&$1+\ln\pi$  \\
\hline
0 & 1 & 1.5772 & 0.6931 & 2.2694 & 2.1447 \\
  & 2 & 0.9248 & 1.2692 & 2.1940 & 2.1447 \\
  & 3 & 0.6475 & 1.5280 & 2.1755 & 2.1447 \\
  & 4 & 0.4698 & 1.6974 & 2.1672 & 2.1447 \\
\hline
1 & 2 & 1.7218 & 1.2692 & 2.9910 & 2.1447 \\
  & 3 & 1.1369 & 1.7697 & 2.9066 & 2.1447 \\
  & 4 & 0.8796 & 1.9736 & 2.8532 & 2.1447 \\
  & 5 & 0.7117 & 2.1089 & 2.8206 & 2.1447 \\
\hline
2 & 3 & 1.7812 & 1.4081 & 3.1893 & 2.1447 \\
  & 4 & 1.2347 & 2.0198 & 3.2545 & 2.1447 \\
  & 5 & 0.9945 & 2.2327 & 3.2272 & 2.1447 \\
  & 6 & 0.8359 & 2.3621 & 3.1980 & 2.1447 \\
\hline
3 & 4 & 1.8110 & 1.5182 & 3.3183 & 2.1447 \\
  & 5 & 1.2928 & 2.1809 & 3.4737 & 2.1447 \\
  & 6 & 1.0662 & 2.4098 & 3.4760 & 2.1447 \\
  & 7 & 0.9157 & 2.5422 & 3.4579 & 2.1447 \\
\end{tabular}
\end{ruledtabular}
\end{table}

\section{Conclusion}
We have studied the information entropies of a class of quantum
systems belonging to the Morse potential. Numerical results for the
position space entropies and momentum space entropies of the Morse
potential are obtained for several potential strengths $\lambda$ and
quantum number $n$. The entropy densities for the above cases were
depicted graphically, for demonstrating the entropy distribution in
the well. It is found that these entropies satisfy the Beckner,
Bialynicki-Birula and Mycielski (BBM) inequality for the Morse
potential.

\section*{Acknowledgements}
This research partially supported by Dokuz Eylul University with
project No: 04.KB.FEN.098, and also was partially supported by the
Scientific and Technological Research Council of Turkey.

\end{document}